\documentstyle[12pt]{article}
\begin{document}
\begin{titlepage}
\title{Scalar Perturbations and Conformal Transformation}
\author{J\'ulio C\'esar Fabris\\
\mbox{\small Departamento de F\'{\i}sica}\\
\mbox{\small Universidade Federal do Esp\'{\i}rito Santo}\\
\mbox{\small Vit\'oria - Esp\'{\i}rito Santo - CEP29060-900}\\
\mbox{\small Brazil}\\
and\\
Jo\"el Tossa\\
\mbox{\small IMSP - Institut de Math\'ematique et Sciences Physiques}\\
\mbox{\small Bo\^{\i}te Postale 613 - Porto Novo}\\
\mbox{\small Benin}}
\maketitle
\begin{abstract}
The non-minimal coupling of gravity to a scalar field can be transformed
into a minimal coupling through a conformal transformation.
We show how to connect the results of a perturbation calculation,
performed
around a Friedmann-Robertson-Walker background solution,
before and after the conformal transformation. We work in the
synchronous gauge, but we discuss the implications of
employing other formalisms.
\end{abstract}
\end{titlepage}
\section{Introduction}
Although the existence of an interaction represented by a scalar field
is strongly restricted by the evidences coming from local gravitational
effects\cite{1}, its cosmological consequences remain object of intensive 
studies. It is possible to imagine many coherent scenarios
where a scalar field may have played a major role in the primordial Universe,
without contradicting the observational data we have today. This is the
case of the Inflationary Scenario, with its many different versions
\cite{2}.
In particular, there was recently a revival of the prototype of
tensor-scalar theories, the Brans-Dicke theory, since it predicts a power-law
behaviour for the scale factor, instead of an exponential one,
avoiding many problems that appear in the traditional inflationary model
\cite{3,4}.
However, in order to avoid contradictions with observation, the 
dimensionless coupling $\omega$ that appears in the Brans-Dicke theory
must be a function of the scalar field itself, so that it can take small
values initially, assuming large values after the inflationary phase,
leading to acceptable values for this parameter today ($\omega > 500$)
\cite{5}.
\par
The possible existence of a scalar field in the primordial universe
leads to the important question of how cosmological perturbations behave
during the period in which such a field play an important role. This has 
been the
object of many studies recently \cite{6,7,8}. However, frequently this 
calculation
is performed in the case where the scalar field is coupled minimally to 
gravity.
In principle, this does not exclude a Brans-Dicke type theory, since this 
theory can be
recast in the form of a minimal coupling through a conformal transformation.
\par
The question we would like to answer here concerns the
equivalence of perturbation
calculations performed in the minimal coupling (so, after the conformal
transformation) with those performed in the non-minimal coupling (before
the conformal transformation). The answer to this question may
be relevant for any kind of
perturbation formalism we employ. Today, we can quote
three main formalisms: the gauge invariant formalism, first proposed by
Bardeen\cite{7,9}; the covariant formalism, proposed by Ellis and Bruni
\cite{10}; finally,
the gauge dependent formalism, which was the first to be used,
in the classic papers of Lifschitz and Khalatnikov\cite{11,12}.
\par
In the case of Bardeen's formalism, the basic variables are generally
constructed in the minimal coupling; so, theories with non-minimal coupling
can be recast in the standard form by employing a conformal transformation.
Anyway, the non-standard variables can be obtained by performing the
inverse conformal transformation since any combination of gauge invariant
variables is also gauge invariant.
However, care must be taken since the solutions in
the Bardeen's formalism are not given in terms of the density contrast,
but in terms of auxiliary quantities. Anyway, even in this formalism,
the actual calculations are performed in a particular frame. Therefore,
we must ask which kind of observer we are considering when we pass
from the non-minimal to minimal coupling, employing however the
same coordinate condition to simplify the calculations.
\par
Concerning the covariant formalism, the fundamental quantity is the
spatial variation of the density contrast, projected in the spatial
section at constant time. It turns out that this quantity is already
conformal invariant. This can be seen as a positive aspect, but we can
also ask what is the real physical meaning of this quantity, since
the original theory does not possess conformal invariance.
\par
The density contrast is, on the other hand, the fundamental quantity
in the Lifschitz-Khalatnikov formalism (seen as the prototype of the
frame-dependent formalism), which can be seen as a positive aspect. We 
will investigate the consequences of solving the perturbative
equations before and after a conformal transformation
when we work with a specific coordinate condition, which
will be the synchronous coordinate condition in our case. This is 
equivalent to ask to which kind of observer we are considering
when we perform the calculation, in a given frame, employing a certain
coordinate condition.
We find that it can be possible to find
the results of the later case from the former ones
if the coordinate condition choice is compatible
with the conformal transformation itself. Under such condition, it is
possible to reobtain the perturbation solutions in the non-minimal coupling
from the perturbation solutions in the minimal coupling.
\par
Here, in order to present an explicit example, we will consider the
case of vacuum solutions. It means that our model will contain just
gravity and a scalar field, without any phenomenological matter
or any other interaction. We will work with the traditional Brans-Dicke
theory, but all the discussion can be extended to any other kind of
tensor-scalar gravity theory.
\par
We organize this paper as follows. In the next section we writte down
the field equations in the case of the Brans-Dicke theory before and
after the conformal transformation and we determine the background solutions,
for a Robertson-Walker metric, showing how to connect them by the inversion
conformal transformation. In third section, we study the scalar perturbations
in both cases, and we determine the analytical solutions when we employ the
synchronous coordinate condition. We show that these solutions can not
be connect by a conformal transformation. In the fourth section, we 
study again the same problem but choosing different coordinate conditions
in the minimal and non-minimal coupling cases,
which are dictated by the conformal transformation, and we show that in 
this case
we can pass from one case to another by such a transformation.
In the fifth section we discuss the question of the residual coordinate
freedom, determining which solutions are physical. Finally, in the last
section, we discuss the results obtained. In the appendix we consider
briefly the case of the tensorial and vectorial modes, which is quite trivial
in the example considered here.
\section{Tensor-Scalar Gravity Theory and the Conformal
Transformation}
The Lagrangian of the the Brans-Dicke theory, which can represent
with quite of generality the tensor-scalar theories, is,
\begin{equation}
\label{nmc}
{\it L} = \sqrt{-g}(\Phi R - \omega\frac{\Phi_{;\rho}\Phi^{;\rho}}{\Phi})
\quad .
\end{equation}
The corresponding field equations are,
\begin{eqnarray}
\label{fea}
R_{\mu\nu} - \frac{1}{2}g_{\mu\nu}R &=& \frac{\omega}{\Phi^2}(\Phi_{;\mu}
\Phi_{;\rho} - \frac{1}{2}g_{\mu\nu}\Phi_{;\rho}\Phi^{;\rho}) +
\frac{1}{\Phi}(\Phi_{;\mu;\nu} - g_{\mu\nu}\Box\Phi) \quad ; \\
\label{feb}
\Box\Phi &=& 0 \quad .
\end{eqnarray}
Here our conventions are $R_{\mu\nu} = \partial_\lambda{\Gamma^\lambda}
_{\mu\nu} - \partial_\nu{\Gamma^\lambda}_{\nu\lambda} +
{\Gamma^\lambda}_{\mu\nu}{\Gamma^\rho}_{\lambda\rho} -
{\Gamma^\lambda}_{\mu\rho}{\Gamma^\rho}_{\nu\rho}$ and
$sig(g) = (+ - - -)$.
Inserting in (\ref{fea},\ref{feb}) the spatially flat Robertson-Walker 
metric,
\begin{equation}
\label{m}
ds^2 = dt^2 - a(t)^2(dx^2 + dy^2 + dz^2) \quad ,
\end{equation}
we find the following equations of motion for $a(t)$ and $\Phi(t)$:
\begin{eqnarray}
3(\frac{\dot a}{a})^2 &=& \frac{\omega}{2}(\frac{\dot\Phi}{\Phi})^2  -
3\frac{\dot a}{a}\frac{\dot\Phi}{\Phi} \quad ; \\
\ddot\Phi + 3\frac{\dot a}{a}\Phi &=& 0 \quad .
\end{eqnarray}
These equations admit the power law solutions,
\begin{eqnarray}
\label{bsa}
a(t) &\propto& t^r \quad , \quad r = \frac{1 + \omega \pm \sqrt{1 +
\frac{2}{3}\omega}}{4 + 3\omega} \quad , \\
\label{bsb}
\Phi(t) &\propto& t^s \quad , \quad s = 1 - 3r \quad .
\end{eqnarray}
\par
If we perform a conformal transformation on the metric $g_{\mu\nu}$
\cite{13}
such that,
\begin{equation}
\label{ct}
g_{\mu\nu} = \Phi^{-1}{\bar g}_{\mu\nu} ,
\end{equation}
we obtain the new Lagrangian:
\begin{equation}
\label{mc}
{\it L} = \sqrt{-\bar g}(\bar R - (\omega + 
\frac{3}{2})\frac{\Phi_{;\rho}\Phi^{;\rho}}
{\Phi^2}) \quad . 
\end{equation}
From (\ref{mc}) we derive the field equations:
\begin{eqnarray}
\label{fec}
\bar R_{\mu\nu} - \frac{1}{2}\bar g_{\mu\nu}\bar R &=& 
\frac{C(\omega)}{\Phi^2}(\Phi_{;\mu}
\Phi_{;\rho} - \frac{1}{2}\bar g_{\mu\nu}\Phi_{;\rho}\Phi^{;\rho}) \quad 
; \\
\label{fed}
\Box\Phi - \frac{\Phi_{;\rho}\Phi^{;\rho}}{\Phi} &=& 0 \quad .
\end{eqnarray}
We have written $C(\omega) = \omega + \frac{3}{2}$.
Inserting in these field equations the metric (\ref{m}), where we note 
now $\tau$ for
the proper time in this new frame, we obtain the following
equations of motion,
\begin{eqnarray}
3(\frac{\dot{\bar a}}{\bar a})^2 &=& 
\frac{C(\omega)}{2}(\frac{\dot\Phi}{\Phi})^2 
\quad ; \\
\ddot\Phi + 3\frac{\dot{\bar a}}{\bar a}\Phi - \frac{\dot\Phi^2}{\Phi} 
&=& 0 \quad .
\end{eqnarray}
which admit the solutions,
\begin{eqnarray}
\label{bsc}
\bar a(t) &\propto& \tau^{\frac{1}{3}} \quad ,\\
\label{bsd}
\Phi(t) &\propto& \tau^n \quad , \quad n = 
\sqrt{\frac{2}{3}}\frac{1}{\sqrt{\frac{2}{3}\omega + 1}} \quad .
\end{eqnarray}
\par
We have the following relations between $\tau$, $t$ and $\bar a(t)$, $a(t)$:
\begin{eqnarray}
\label{cra}
t &=& \tau^{-\frac{n}{2} + 1} \quad , \\
\label{crb}
a &=& \Phi^{-\frac{1}{2}}\bar a
\end{eqnarray}
So, using (\ref{bsc},\ref{bsd}) and (\ref{cra},\ref{crb}),
we reobtain (\ref{bsa},\ref{bsb}).
\section{Scalar Perturbations}
Now, we will evaluate an analytic expression for the scalar perturbations
around the background solutions found above. We will calculate them both
in the case of minimal coupling (which is also called Einstein's frame),
as in the case of the non-minimal coupling (also called Jordan's frame).
All these calculations will be performed in the synchronous gauge
\cite{11,12} $h_{\mu0} = 0$. 
\par
In this computation, we will consider that the perturbed quantities
behave spatially as plane waves,
\begin{equation}
\delta(t,\vec x) = \delta(t)e^{i\vec q.\vec x}
\end{equation}
where $\vec q$ is the wavenumber of the perturbation. So, all laplacian
operatores acting on the three-dimensional spatial section can be
replaced by $-q^2$.
\subsection{Perturbations in the Minimal Coupling}
In what follows, we will suppresse the bars in order to simplify the
notation.
We write the field equations as,
\begin{eqnarray}
R_{\mu\nu} &=& C(\omega)\frac{\Phi_\mu\Phi_\nu}{\Phi^2} \quad , \\
\Box\Phi - \frac{\Phi_{;\rho}\Phi^{;\rho}}{\Phi} &=& 0 \quad .
\end{eqnarray}
We introduce now the perturbed quantities,
\begin{eqnarray}
\tilde g_{\mu\nu} &=& g_{\mu\nu} + h_{\mu\nu} \quad , \\
\tilde\Phi &=& \Phi + \delta\Phi \quad ,
\end{eqnarray}
where $g_{\mu\nu}$ and $\Phi$ are the background solutions (\ref{bsc},
\ref{bsd}) and $h_{\mu\nu}$ and $\delta\Phi$ are small perturbations
around them.
\par
Fixing the synchronous coordinate condition $h_{0\mu} = 0$, and defining
$h = \frac{1}{2}\frac{h_{kk}}{a^2}$ and $\lambda = 
\frac{\delta\Phi}{\Phi}$, we obtain, after a quite long
calculation, the following expressions linking $h$ and $\lambda$:
\begin{eqnarray}
\label{pea}
\ddot h + 2(\frac{\dot a}{a})\dot h &=& 
2C(\omega)\frac{\dot\Phi}{\Phi}\dot\lambda
\quad ; \\
\label{peb}
\ddot\lambda + 3\frac{\dot a}{a}\dot\lambda 
+ \frac{q^2}{a^2}\lambda &=& \frac{\dot\Phi}{\Phi}
\dot h \quad .
\end{eqnarray}
The dots means derivative with respect to the proper time in the
Einstein's frame, $\tau$.
The integration of this equation is not very difficult\cite{14}. First of 
all,
we change for the conformal time, $\eta = q\int{\frac{d\tau}{a}}$. Using
(\ref{peb}), we eliminate the terms in $h'$ and $h''$ in (\ref{pea}),
where the primes mean derivative with respect to the conformal time.
So, we obtain a third order equation for $\lambda$:
\begin{equation}
\label{fea1}
\lambda''' + \frac{5}{2}\frac{\lambda''}{\eta} + 
\biggr(1 - \frac{5}{2}\frac{1}{\eta^2}\biggl)\lambda' + \frac{3}{2}
\frac{\lambda}{\eta} = 0 \quad .
\end{equation}
To derive these equation we have also expressed the solutions
(\ref{bsc},\ref{bsd}) in terms of the conformal time. To solve (\ref{fea1}),
we proceed as follows. We write $\lambda = \eta^{-\frac{3}{2}}\Psi$. With 
this
definition, the equation (\ref{fea1}) reduces to a second order equation.
Then, writting $\Psi' = \eta^{\frac{3}{2}}\gamma$, we reduce this equation
to a Bessel's equation of first kind.
The final solution, now in terms of the proper time $\tau$, is:
\begin{equation}
\label{fsa}
\lambda = 
\frac{1}{\tau}\biggr(\int{\tau^{-\frac{2}{3}}[c_1J_1(\tau^{\frac{2}{3}})
+ c_2N_1(\tau^{\frac{2}{3}})]d\tau} + c_3\biggl) \quad ,
\end{equation}
where $J_1$ and $N_1$ are Bessel's and Neumann's functions of first kind,
and $c_1$, $c_2$ and $c_3$ are constants. As we will see later, the solution
represented by the integration constant $c_3$ can be eliminated by a
coordinate transformation.
\par
The solution for $h$ can be obtained from (\ref{fsa}) and (\ref{peb}).
The result is:
\begin{eqnarray}
\label{fsb}
nh &=& \int_0^\tau{v^{-2}\biggr(\int_0^v{(u^{\frac{2}{3}} + 
\frac{2}{3}u^{-\frac{2}{3}})[c_1J_1(u^{\frac{2}{3}}) +
c_2N_1(u^{\frac{2}{3}})]du} + c_3\biggl)dv}  + \nonumber \\
& & + \int_0^\tau{\bigg(\frac{2}{3}v[c_1J'_1(v^{\frac{2}{3}}) 
+ c_2N'_1(v^{\frac{2}{3}})] - \frac{1}{3}[c_1J_1(v^{\frac{2}{3}}) 
+ c_2N_1(v^{\frac{2}{3}})]\biggl)dv} + \nonumber \\
& & + c_4 \quad .
\end{eqnarray}
\subsection{Perturbations in the Non-Minimal Coupling}
The field equations (\ref{fea},\ref{feb}) can be rewritten as
\begin{eqnarray}
R_{\mu\nu} &=& \frac{\omega}{\Phi^2}\Phi_\mu\Phi_\nu + 
\frac{1}{\Phi}\Phi_{;\mu;\nu} \quad , \\
\Box\Phi &=& 0 \quad .
\end{eqnarray}
We will proceed exactly as before in order to calcule the perturbed 
equations and solve them. But, in order to distinguish the calculations
performed before the conformal transformation from those performed after
the conformal transformation, we note now $\tau$ as the proper time,
$H = \frac{1}{2}\frac{h_{kk}}{a^2}$ where now all quantities are calculeted
in the Jordan's frame, and $a$ is now given by (\ref{bsa}).
\par
Introducing the calculations as in the preceding case,
we find the following differential equations linking $H$ and $\lambda$:
\begin{eqnarray}
\label{eqfa}
\ddot H + 2\frac{\dot a}{a}\dot H &=& \ddot\lambda +
2(1 + \omega)\frac{\dot\Phi}{\Phi}\dot\lambda \quad ,\\
\label{eqf}
\ddot\lambda + (3\frac{\dot a}{a} + 2\frac{\dot\Phi}{\Phi})\dot\lambda +
(\frac{q}{a})^2\lambda &=& \frac{\dot\Phi}{\Phi}\dot H \quad .
\end{eqnarray}
Now the dots means derivatives with respect to the proper time in the
Jordan's frame.
We solve these coupled equations in the same way as before.
The final results are\cite{15}:
\begin{eqnarray}
\label{fsc}
\lambda &=& \frac{1}{t}\biggr(\int_0^t{u^{1-r}[b_1J_1(qu^{1-r})  
b_2N_1(qu^{1-r})]}du + b_3\biggl) \quad ; \\
\label{fsd}
(1-3r)H &=& \int_0^t{\biggr([\frac{3r}{1-r}u^{r-2} +
(1-r)u^{-r}]}\int_0^u{v^{1-r}[b_1J_1(qv^{1-r})}\nonumber \\ 
& & + b_2N_1(qv^{1-r})]dv +
b_3\biggl)du + \nonumber \\
& & + \int_0^t{\biggr((1 - 4r)[b_1J_1(qu^{1-r}) +
b_2N_1(qu^{1-r})] +} \nonumber \\
& & + (1-3r)[b_1J'_1(qu^{1-r}) +
b_2N'_1(qu^{1-r})]\biggl)du \quad .
\end{eqnarray}
\section{Connecting the Perturbations}
It is possible to verify that the solutions (\ref{fsa},\ref{fsb}) do not
correspond to the solutions (\ref{fsc},\ref{fsd}). The proper time $t$ in the
Jordan's frame is related to the proper time $\tau$ in the Einstein's
frame by the relation (\ref{cra}).
Inserting this relation in
(\ref{fsa},\ref{fsb}) we see that they correspond to a solution
which is different
from (\ref{fsc},\ref{fsd}).
\par
This problem arises from our choice of coordinate conditions. To the
obtain the above solutions we have imposed the synchronous coordinate
condition in the Jordan's frame and in the Einstein's frame. But, this
is not compatible with the conformal transformation. Considering the
conformal transformation (\ref{ct}), we have for the perturbations,
\begin{equation}
H_{\mu\nu} = - \frac{\lambda}{\Phi}g_{\mu\nu} + \frac{1}{\Phi}h_{\mu\nu}
\end{equation}
where $H_{\mu\nu}$ is the metric perturbation in the Jordan's frame and
$g_{\mu\nu}$ and $h_{\mu\nu}$ are the metric and its perturbation in
the Einstein's frame respectively. So, if we impose the synchronous
coordinate condition in the Jordan's frame, in order to retain a
coherence with the conformal transformation,
we find that in the Einstein's frame, we must impose the conditions(which,
for simplicity, we call conformal frame, in contrast with the
synchronous frame),
\begin{eqnarray}
h_{00} &=& \lambda \quad , \\
h_{i0} &=& 0 \quad . 
\end{eqnarray}
With these conditions, we find the following equations linking $h$, defined
as before, and $\lambda$:
\begin{eqnarray}
\label{pee}
\ddot h + 2(\frac{\dot{\bar a}}{\bar a})\dot h &=& 
(2C(\omega)\frac{\dot\Phi}{\Phi} -
\frac{3}{2}\frac{\dot{\bar a}}{\bar a})\dot\lambda + \frac{1}{2}
(\frac{q}{\bar a})^2\lambda
\quad ; \\
\label{pef}
\ddot\lambda + (3\frac{\dot{\bar a}}{\bar a} - \frac{1}{2}\frac{\dot\Phi}
{\Phi})\dot\lambda 
+ (\frac{q}{\bar a})^2\lambda &=& \frac{\dot\Phi}{\Phi}
\dot h \quad .
\end{eqnarray}
\par
The dots here means derivative with respect to $\tau$. To solve this 
equation, we follow the same procedure as before: we pass to the 
conformal time, and
with the help of (\ref{pef}), we eliminate $\dot h$ and $\ddot h$ in
(\ref{pee}). We obtain a third order equation for $\lambda$, which, with the
help of the background solutions (\ref{bsc},\ref{bsd}), can be written as,
\begin{equation}
\lambda''' + (s + 1)\frac{\lambda''}{x} + \biggr(1 - (\frac{1 + s}{x^2})
\biggl)\lambda' + s\frac{\lambda}{x} = 0 \quad .
\end{equation}
Defining $\lambda = x^{-s}\omega$,
where $s = \frac{3}{2}(1 - \frac{n}{2})$. we obtain
a second order differential equation. Writting
$\omega' = x^s\gamma$, this equation takes the form of a Bessel's equation
of the first kind. The final solution for $\lambda$ in terms of the proper
time $t$ is:
\begin{equation}
\label{fs}
\lambda = \frac{1}{\tau^{\frac{2s}{3}}}\biggr(\int_0^\tau{u^{\frac{2s-1}{3}}
[c_1J_1(qu^{\frac{2}{3}}) + c_2N_1(qu^{\frac{2}{3}})]du} + c_3\biggl)
\quad .
\end{equation}
\par
Using (\ref{cra}) we can now obtain from (\ref{fs}) the solution 
(\ref{fsc}). We can also obtain (\ref{fsd}), but it is more easy to
show that the equation (\ref{pee}) reduces to (\ref{eqf}).
This can be done using the relations,
\begin{eqnarray}
H &=& \frac{3}{2}\lambda + h \quad , \\
\bar a &=& \Phi^{-\frac{1}{2}}a \quad .
\end{eqnarray}
Under these transformations, which are induced by the conformal 
transformation
itself, we can reobtain the perturbed equations in the Jordan's frame from
the perturbed equations in the Einstein's frame and vice-versa.
\section{Residual Coordinate Freedom}
In spite of the fact that we have already fixed a coordinate condition,
it remains a residual coordinate freedom so that some of the modes
found above can have no physical meaning\cite{16}. We will consider this 
problem
both in the case of the no-minimal coupling, with the synchronous coordinate
condition, as in the case of the minimal coupling, with the coordinate
conditions that is compatible with the conformal transformation, that
we will call henceforth conformal frame.
\subsection{Residual Coordinate Freedom in the Synchronous Frame}
By an infinitesimal coordinate transformation of the kind,
\begin{equation}
x^\mu \rightarrow x^\mu + \chi^\mu
\end{equation}
the perturbed metric changes as,
\begin{equation}
H_{\mu\nu} \rightarrow H_{\mu\nu} + \chi_{\mu;\nu} + \chi_{\nu;\mu}
\quad .
\end{equation}
Defining $\chi^0 = \chi$ and $\chi^i = \theta^{,i}$, employing the
synchronous coordinate condition and imposing that this transformation
preserves the coordinate condition, we obtain, that
\begin{eqnarray}
H &\rightarrow& H - \theta_{,k,k} - 3\frac{\dot a}{a}\chi \quad , \\
\theta &=& \chi\int{\frac{dt}{a^2}} \quad .
\end{eqnarray}
where $H = \frac{1}{2}\frac{H_{kk}}{a^2}$ and $\theta$ and $\chi$ are 
time-independent functions.
A solution for $h$ is non-physical if it can be written as,
\begin{equation}
H = \theta_{,k,k} + 3\frac{\dot a}{a}\chi \quad .
\end{equation}
Inserting this in (\ref{eqfa},\ref{eqf}), we obtain the foloowing
solution for $\lambda$

\begin{equation}
\lambda = \frac{(1-3r)\chi}{t} \quad .
\end{equation}
The mode represented by the integration constant $b_3$ in
(\ref{fsc}) can be eliminated by a coordinate transformation and has
no physical meaning.
\subsection{Residual Coordinate Freedom in the Conformal Frame}
If we employ the same coordinate transformation, with respect to the
coordinate condition employed in the last section in the
conformal frame, we have,
\begin{eqnarray}
h_{00} \rightarrow h_{00} + 2\dot\chi_0 \quad , \\
h_{0i} \rightarrow h_{0i} + \chi_{0;i} + \chi_{i;0} \quad ,\\
h_{ij} \rightarrow h_{ij} + \chi_{i;j} + \chi_{j;i}  \quad .
\end{eqnarray}
Remembering that $h_{00} = \lambda$, and that a gauge solution must
made zero by an infinitesimal coordinate transformation, we obtain the 
following solution
for $h = \frac{1}{2}\frac{h_{kk}}{{\bar a}^2}$:
\begin{equation}
h = \int{\frac{\chi_{,k,k}}{{\bar a}^2}dt} + 3\frac{\dot{\bar a}}{\bar 
a}\chi \quad ,
\end{equation}
where $\dot\chi = - \frac{\lambda}{2}$.
\par
Inserting this in the equation (\ref{pee},\ref{pef}) we can determine a 
general solution
for $\chi$:
\begin{equation}
\chi = \chi_0t^{\frac{n}{2}} \quad .
\end{equation}
\par
In the same way, returning to $\lambda$, we can see that the solution
represented by $c_3$ in (\ref{fsa}) can also be eliminated by a 
coordinate transformation, and it has also no physical meaning.
All modes calculated in one frame are directly related with the
modes calculated in the other frame.
\section{Conclusion}
A non-minimal coupling between gravity and a scalar field, like in the
Brans-Dicke theory, can be put generally in the form of a minimal 
coupling through a conformal transformation. The motivations to do this 
can be
twofold: we can perform such a transformation for technical reasons,
looking for a frame in which the analysis of the problem can be more
transparent; or because the physical content of the theory is in the
metric field of the minimal coupling and not of the non-minimal coupling,
or vice-versa.
\par
Regarding just the background solution, this question is quite trivial,
since we can easily pass from one frame to the other. We have showed here
that, on the other hand, at the perturbative level, there is no clear
equivalence when we perform the calculation in one frame or in the other.
We have exemplified this problem in the context of the Brans-Dicke theory.
This choice was made due to the great generality of the Brans-Dicke
theory, and in fact it can easily be transposed to other tensor-scalar 
theories.
\par
We have shown that, in order to permit an equivalence of the calculation
in the Einstein's frame with respect to the Jordan's frame,
assuring that we are talking about the same class of observer,
the coordinate
conditions employed must respect the conformal transformation itself.
If we choose a coordinate condition in one frame, we must choose another
coordinate condition in the other frame, which is related to the first
one by a conformal transformation. If we do this properly, then it is
possible to pass from one frame to the other coherently at the perturbative
level. In particular,
physical modes are transported into physical ones, and the non-physical
modes are also transported into non-physical modes, as it could be
expected.
\par
We have performed our analysis using a particular coordinate condition.
But this problem can be also analysed in the case of the so-called
gauge invariant formalism\cite{7}.  There, from the point of view of
the construction of gauge invariant functions, the using of
a conformal transformation brings no difficulty since any
combination of a gauge invariant quantity is also gauge invariant. However,
the question of what observers we are talking about when we change
from the Jordan's frame to Eisntein's frame remains\cite{17}.
\newline
\leftline{\bf Acknowledgements} We would like to thank the hospitality of 
{\it International Centre for Theoretical Physics}, Trieste, Italy,
and of the {\it Laboratoire de Gravitation et Cosmologie Relativistes},
Paris, France, during the elaboration of this work. J.C.F. thanks also the
{\it CNPq}, Brazil, for financial support. We thank also
Nelson Pinto Neto for a carefull reading of this text.
\section*{Appendix: The Tensorial and Vectorial Modes}
We will consider now how the tensorial and vectorial modes
behave under a conformal transformation. In a matter of fact,
when we consider the case of gravity coupled to a scalar field,
the solutions are quite trivial for this case: the scalar field
has no direct influence on the tensorial and vectorial modes, as we
shall see.
\par
A second order tensor can be decomposed into a pure tensorial
term, a vectorial term and two scalar terms. The decomposition
has the form,
\begin{equation}
h_{ij} = T_{ij} + V_{(i;j)} + S_{;i;j} + \tilde S\gamma_{ij} \quad ,
\end{equation}
where $T_{ij}$ is a tracelless and transverse tensorial term,
$V_i$ is a divergence free term and $S$, $\tilde S$ are two different
scalar modes. The derivatives are with respect to the homogenous space
section and $\gamma_{ij}$ is the metric on this homogenous space.
This is equivalent to a division into spin components $2$, $1$ and $0$.
\par
To perform the calculations we calculate the perturbed components of
the Ricci tensor $\delta R_{\mu\nu}$, and decompose all the functions
into tensorial, vectorial and scalar eigenfunctions.
Taking the components of the tensorial mode only, considering the
properties described above, we obtain the following equation:
\begin{equation}
\ddot T_{ij} - \frac{\dot a}{a}\dot T_{ij} +((\frac{q}{a})^2 +
4(\frac{\dot a}{a}^2)T_ij = 0 \quad .
\end{equation}
Note that the influence of the scalar field enters only in the
behaviour of the scale factor.
The solution for this equation is,
\begin{equation}
T_{ij} = x^p(c_1J_\nu(x) + c_2J_{-\nu}(x))Q_{ij}
\end{equation}
where $x$ is the conformal time, $c_1$ and$c_2$ are constants,
$p =r +\frac{1}{2} + r$, $a \propto x^r$ and $Q_{ij}$ is the tensorial
eigenfunction. The result above is valid
both in synchronous ($h_{0\mu} = 0$) or conformal ($h_{0i} = 0, h_{00} =
\lambda$) gauge .
\par
Concerning the vectorial modes, we have the equation
\begin{equation}
\dot F_i - 2\frac{\dot a}{a}F_i = 0 \quad ,
\end{equation}
where $F_i = \nabla V_i$, where the operator $\nabla = 
\gamma^{ij}\partial_i\partial_j$ is defined in the
homogenous space. The solution is
\begin{equation}
F_i = ca^2Q_i
\end{equation}
$c$ being a constant, $a$ the scale factor and $Q_i$ a vectorial
eigenfunction. Again this result is valid in both gauges, the difference
of working in the Einstein's or Jordan's frame comming from the
behaviour of the scale factor.

\end{document}